\documentclass[letterpaper, 10 pt, journal]{IEEEtran}
\usepackage{amsmath,amsfonts}
\usepackage{algorithmic}
\usepackage{algorithm}
\usepackage{array}
\usepackage[caption=false,font=normalsize,labelfont=sf,textfont=sf]{subfig}
\usepackage{textcomp}
\usepackage{stfloats}
\usepackage{url}
\usepackage{booktabs}
\usepackage{verbatim}
\usepackage{graphicx}
\usepackage{cite}
\usepackage{makecell}
\usepackage{multirow}
\begin{document}

\title{Zero-Shot, Safe and Time-Efficient UAV Navigation via Potential-Based Reward Shaping, Control Lyapunov and Barrier Functions}

\author{Ashik Abrar Naeem and Mohammad Ariful Haque}

\maketitle

\begin{abstract}
Autonomous navigation and obstacle avoidance remain a core challenge of modern Unmanned Aerial Vehicles (UAVs). While traditional control methods struggle with the complexity and variability of the environment, reinforcement learning (RL) enables UAVs to learn adaptive behaviors through interaction with the environment. Existing research with RL prioritizes the mission success at the expense of mission time and safety of UAVs. This study integrates Potential Based Reward Shaping (PBRS) with Control Lyapunov Functions (CLF) and Control Barrier Functions (CBF) to simultaneously optimize mission time and ensure formal safety guarantees. An RL model is trained in a generalized simple environment, then used in complex scenarios incorporating a CLF-CBF-QP filter without further training. Experimental results in simulated environments demonstrate a significant reduction in mission time and outstanding performance in complex environment.
\end{abstract}


\section{Introduction}
\IEEEPARstart{N}{avigation} remains a fundamental challenge for Unmanned Aerial Vehicles (UAVs), underpinning a wide array of emerging applications \cite{li2026review, ji2025uav, zhang2026uav}.
While traditional control methods have laid the groundwork for autonomy, recent efforts have shifted toward Reinforcement Learning (RL) \cite{kendoul2012survey, rezwan2022artificial, aburaya2024review}. The appeal of RL lies in its ability to derive control policies through environmental interaction without requiring an exhaustive aggregate description of the workspace. However, the performance of these agents is inextricably linked to the quality of the reward function - a component that remains notoriously difficult to design.

Current research primarily addresses sample inefficiency and task performance through specialized architectures or reward schemes. For instance, researchers have explored structural enhancements such as Multiple Pool TD3 for target tracking \cite{xu2022autonomous}, Transformer blocks to mitigate information loss \cite{jiang2024autonomous}, and Quantum-inspired experience replay \cite{ma2023target}. Despite these innovations, the underlying reward functions are often designed heuristically. While Inverse Reinforcement Learning can recover rewards from expert data \cite{chen2026learning}, it is constrained by the scarcity of high-quality demonstrations.

Existing reward structures have several limitations. For instance, sparse threshold-based reward in \cite{wang2020deep} may improve success rates, but it provides no informative feedback for time-optimal behavior. The augmented backward rewards in \cite{chansuparp2022novel} require intensive computation through Convolutional Neural Networks. Most critically, many frameworks rely on empirically tuned weights to "force" desired behaviors \cite{tsourveloudis2025uav}. This manual tuning is not only labor-intensive but also brittle; if a new constraint such as mission time is introduced, the entire reward function must be redesigned. To overcome this, we employ Potential-Based Reward Shaping (PBRS), inspired by \cite{ng1999policy}, which allows us to integrate mission-time constraints while mathematically guaranteeing the preservation of the optimal policy.

A secondary, yet equally vital, concern is the inherent instability of RL policies. To provide the formal safety guarantees required for real-world deployment, researchers have integrated Control Lyapunov Functions (CLFs) for stability and Control Barrier Functions (CBFs) for safety. Recent work has utilized High-Order CBFs as explicit safety filters for target interception \cite{peng2023design} and hierarchical frameworks involving probabilistic enumeration to identify unsafe regions offline \cite{marzari2025designing}. Others have used Deep Q-Networks to generate high-level commands for low-level CLF-CBF-QP controllers \cite{chen2025collision}, or incorporated differentiable safety layers directly into the policy gradient updates \cite{emam2022safe}. Despite these advancements, a significant limitation persists: most existing frameworks train the RL agent within the CLF-CBF constraints \cite{xia2025multi}. This coupling makes the policy environment-dependent; if the scenario changes, the agent typically requires retraining. This paper proposes a framework that decouples task-learning from safety-enforcement to achieve zero-shot transfer. The core contributions are:

Time-Efficient Navigation: By incorporating PBRS into the reward design, we significantly reduce mission time without the "trial-and-error" weight tuning common in previous studies.

Decoupled Safety Filter: We train the agent in a generalized, simple environment and wrap it in a CLF-CBF-QP filter during deployment. Unlike prior work that requires retraining for new scenarios, our approach demonstrates outstanding performance in complex, unseen environments without further parameter updates.

\section{Preliminaries}

To establish the theoretical foundation for the proposed UAV navigation framework, this section reviews three core concepts. First, Potential-Based Reward Shaping (PBRS)\cite{ng1999policy} is introduced, which plays the critical role of accelerating the learning process without altering the optimal policy. Second, Control Lyapunov Functions (CLFs)\cite{slotine1991applied} are described, these are utilized within the deployment filter to provide formal stability guarantees, ensuring the UAV continuously and effectively navigates toward its target. Finally, Control Barrier Functions (CBFs)\cite{ames2019control} are discussed, which serve as a rigorous mathematical safety boundary to render the operational space forward invariant, thereby providing formal safety guarantees against collisions. Together, these principles form the backbone of the proposed methodology.

\subsection{Potential-Based Reward Shaping}
Potential-Based Reward Shaping (PBRS) is employed to accelerate the learning process without altering the optimal policy of the agent. The shaped reward $R'$ is defined as
\begin{equation*}
    R'(s_t, a_t, s_{t+1}) = R(s_t, a_t, s_{t+1}) + F(s_t, s_{t+1}),
\end{equation*}
where $s$ is the state and $a$ is the action, $R$ is the original reward and $F$ is the shaping function. $F$ is derived from a potential function $\Phi : \mathcal{S} \to \mathbb{R}$
\begin{equation}
\label{eq:shaping_rewards}
    F(s_t, s_{t+1}) = \gamma \Phi(s_{t+1}) - \Phi(s_t).
\end{equation}
The value function under the shaped reward, denoted as $V'^\pi(s)$, can be derived by expanding the expectation of the discounted sum of shaped rewards
\begin{equation*}
    V'^\pi(s) = \mathbb{E} \left[ \sum_{t=0}^{\infty} \gamma^t R'(s_t, a_t, s_{t+1}) \mid s_0 = s, \pi \right].
\end{equation*}
By substituting the shaping term the relationship between the original and shaped value functions simplifies to
\begin{equation*}
    V'^\pi(s) = V^\pi(s) + \Phi(s).
\end{equation*}
Since $\Phi(s)$ acts as a constant offset for any given state, it does not affect the relative ordering of action values under the policy $\pi$. This property ensures that the optimal policy $\pi^*$ remains the same.

\subsection{Control Lyapunov Function}
To ensure the UAV effectively navigates toward a target state, a Control Lyapunov Function is utilized to provide formal stability guarantees.

Consider a nonlinear control affine system of the form
\begin{equation}
\label{eq:nonlinear_system}
    \dot{x} = f(x) + g(x)u,
\end{equation}
where $x \in D \subset \mathbb{R}^n$ is the system state, $u \in U \subset \mathbb{R}^m$ is the control input, and $f: D \to \mathbb{R}^n$ and $g: D \to \mathbb{R}^{n \times m}$ are locally Lipschitz continuous functions representing the drift and input vector fields, respectively. A continuously differentiable function $V: D \to \mathbb{R}_{\geq 0}$ is defined as a Control Lyapunov Function if it is positive semidefinite and there exists a control input $u$ such that the derivative of $V$ along the system trajectories satisfies
\begin{equation}
\label{eq:clf_condition}
    \inf_{u \in U} [L_f V(x) + L_g V(x)u] \leq -\gamma(V(x)),
\end{equation}

where $L_f V(x) = \nabla V \cdot f(x)$ and $L_g V(x) = \nabla V \cdot g(x)$ represent the Lie derivatives of $V$ along the vector fields $f$ and $g$, respectively. The term $\gamma$ denotes a class $\mathcal{K}$ function, which is strictly monotonic and satisfies $\gamma(0)=0$.

The existence of a CLF implies the existence of a set of stabilizing controllers for every point $x \in D$, defined as
\begin{equation}
\label{eq:clf_controller}
    K_{\text{clf}}(x) := \{u \in U : L_f V(x) + L_g V(x)u \leq -\gamma(V(x))\}.
\end{equation}

Any Lipschitz continuous feedback controller $u(x) \in K_{\text{clf}}(x)$ is guaranteed to asymptotically stabilize the system to the equilibrium point.

\subsection{Control Barrier Function}
To provide guarantee that the UAV remains inside a safe region, a Control Barrier Function is utilized.
Safety is framed as keeping the system state $x$ within a predefined safe set $\mathcal{C}$. $\mathcal{C}$ is defined as the superlevel set of a continuously differentiable function $h: D \subset \mathbb{R}^n \to \mathbb{R}$
\begin{equation}
\label{eq:safe_set}
    \begin{aligned}
        \mathcal{C} &= \{x \in D \subset \mathbb{R}^n : h(x) \geq 0\}, \\
        \partial\mathcal{C} &= \{x \in D \subset \mathbb{R}^n : h(x) = 0\}, \\
        \text{Int}(\mathcal{C}) &= \{x \in D \subset \mathbb{R}^n : h(x) > 0\}.
    \end{aligned}
\end{equation}
The set $\mathcal{C}$ is considered forward invariant if for any initial condition $x_0 \in \mathcal{C}$, the trajectory $x(t)$ remains within $\mathcal{C}$ for all $t \geq 0$.

The function $h$ is a Control Barrier Function if there exists an extended class $\mathcal{K}_\infty$ function $\alpha$ with $\alpha(0) = 0$ that satisfies
\begin{equation}
\label{cbf_condition}
    \sup_{u \in U} [L_f h(x) + L_g h(x)u] \geq -\alpha(h(x)),
\end{equation}

for all $x \in D$.

The set of all safe controllers values is defined as
\begin{equation}
\label{eq:cbf_controller}
    K_{\text{cbf}}(x) = \{u \in U : L_f h(x) + L_g h(x)u + \alpha(h(x)) \geq 0\}.
\end{equation}
Any Lipschitz continuous controller $u(x) \in K_{\text{cbf}}(x)$ renders the set $\mathcal{C}$ forward invariant.

\section{Methodology}
\subsection{Problem Statement}
\begin{figure}[h!] 
    \centering
    \includegraphics[width=1.0\columnwidth]{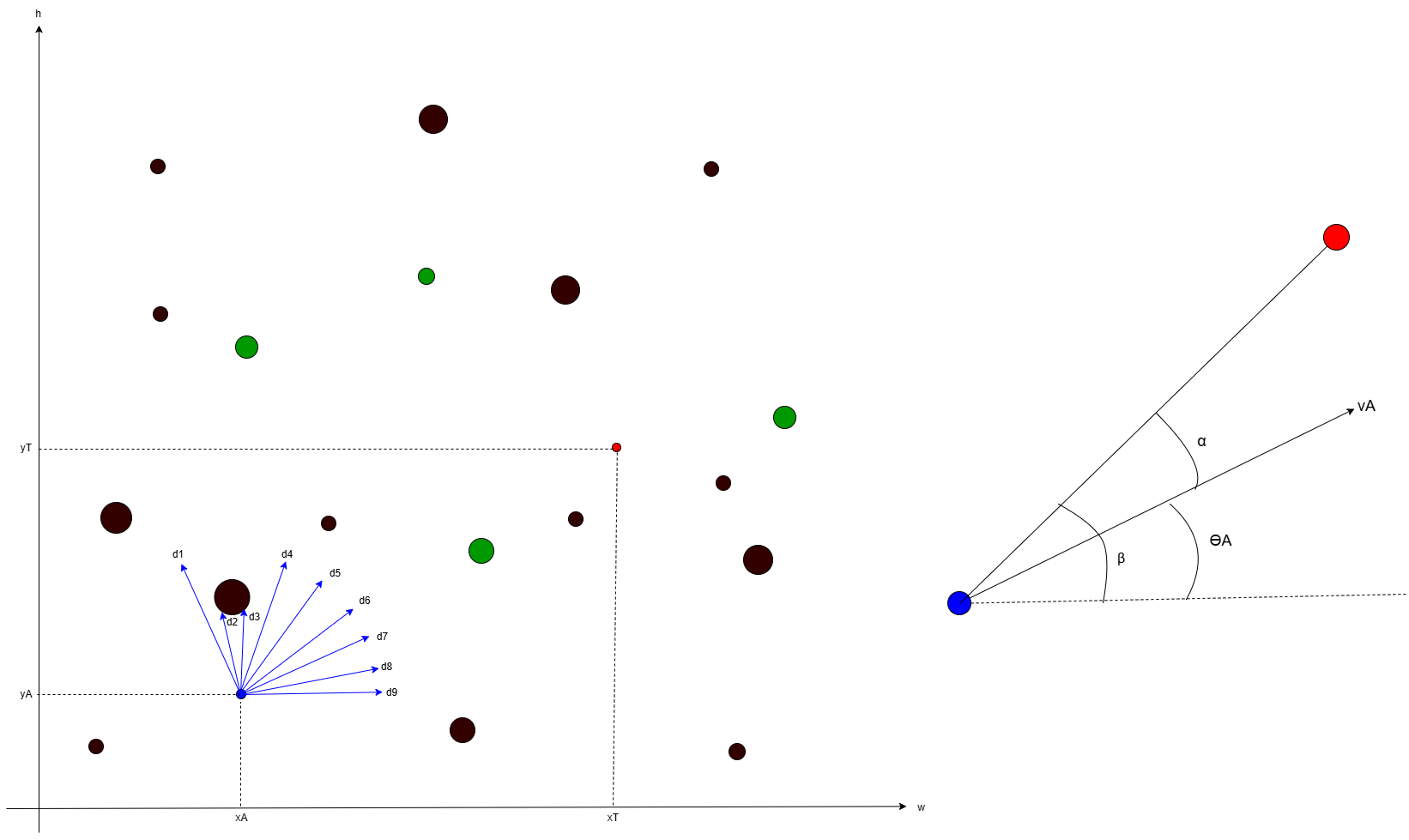} 
    \caption{Schematic diagram of the UAV environment. The agent UAV (blue) pursues the destinaiton (red) while avoiding static (black) and dynamic (green) obstacles. The dynamics obstacles moves in straight line unless they encounter with the environment boundary or another obstacle. The UAV is equiped with onboard sensors like LiDAR that emits 9 rays over a $120^{\circ}$ forward arc to measure the obstacle distance.}
    \label{fig:setup}
\end{figure}

In order to mathematically deal with the UAV navigation problem, the UAV is treated as a sphere flying at a constant altitude. The UAV is operated within a bounded rectangular environment of dimensions $w \times h$, that contains $N_{\text{sta}}$ static and $N_{\text{dyn}}$ dynamic spherical obstacles with radius ranging from $r_{\text{min}}$ to $r_{\text{max}}$. The schematic diagram of the setup is illustrated in Fig.~\ref{fig:setup}. It is assumed that the UAV has the coordinates of the destination point via mechanisms such as GPS. The objective of the UAV is to navigate toward the destination while avoiding obstacles and maintaining a threshold velocity for faster tracking.

\subsection{Solution Approach}
To solve the navigation problem the UAV is treated as an agent in a Markov Decision Process. To obtain the optimal policy, a Multiple Pool augmented TD3 algorithm inspired by \cite{xu2022autonomous} is utilized. The Observation Space, Action Space, Obstacle Kinematics and Reward Functions are described below

\subsubsection{Observation Space}

The agent relies on some onboard sensors for observing the environment. The observation vector $s_t$ at time $t$ is a 18-dimensional normalized representation, define as follows

The UAV's own position, velocity, and heading are normalized and given by
\begin{equation*}
    s_1^t = \frac{x_A^t}{w}, \quad s_2^t = \frac{y_A^t}{h}, \quad s_3^t = \frac{\theta_A^t \bmod 2\pi}{\pi}, \quad s_4^t = \frac{v_A^t}{v_{\max}^A}
\end{equation*}
where $(x_A^t, y_A^t)$ are the 2D coordinates, $\theta_A^t \in [-\pi, \pi]$ is the heading angle, $v_A^t \in [0, v_{\max}^A]$ is the velocity magnitude. The destination position is similarly normalized
\begin{equation*}
    s_5^t = \frac{x_T^t}{w}, \quad s_6^t = \frac{y_T^t}{h}, \quad 
\end{equation*}
where $(x_T^t, y_T^t)$ are the destination's coordinates. The UAV emits 9 rays over a 120-degree forward arc (from $-\pi/3$ to $\pi/3$ relative to $\theta_A^t$), each returning the distance to the nearest obstacle, up to a maximum sensing range $r_{\text{cap}}$. These are encoded as
\begin{equation*}
    s_j^t = \frac{d_i}{r_{\text{cap}}}, \quad i = 1, 2, \ldots, 9, \quad j = 7, 8, \ldots, 15
\end{equation*}
where $d_i \in [0, r_{\text{cap}}]$ is the distance in the $i$-th ray direction. The relative position and orientations of the UAV to the destination are normalized and included in the observation space
\begin{equation*}
    s_{16}^t = \frac{\beta}{\pi}, \quad s_{17}^t = \frac{d_t}{\sqrt{w^2 + h^2}}, \quad s_{18}^t = \frac{\alpha}{\pi}
\end{equation*}
where,
\begin{equation*}
    d_t = \sqrt{(x_T^t - x_A^t)^2 + (y_T^t - y_A^t)^2}
\end{equation*}
is the Euclidean distance, 
\begin{equation*}
    \beta = \arctan(\frac{y_T^t - y_A^t}{x_T^t - x_A^t})
\end{equation*}
is the relative angle to the horizontal, and 
\begin{equation*}
    \alpha = \arccos(\frac{(x_T^t - x_A^t)\cos(\theta_A^t) + (y_T^t - y_A^t)\sin(\theta_A^t)}{d_t})
\end{equation*}
is the alignment angle between the UAV's heading and the UAV-destination line. The angles are shown in Fig.~\ref{fig:setup}.

\subsubsection{Action Space}
The action space of the UAV is designed to enable smooth and controlled motion. The action $a_t \in A$ at time $t$ is a continuous two-dimensional vector $[\lambda_v, \lambda_\theta]$, where $\lambda_v, \lambda_\theta \in [-1, 1]$ represent fractional changes in velocity and heading angle, respectively. The velocity magnitude $v_A^t$ is updated based on $\lambda_v$, with a maximum change of $\Delta v_{\max}$ per time step, and is capped at a maximum flight speed $v_{\max}^A$
\begin{equation*}
    v_A^{t+1} = \text{clip}(v_A^t + \lambda_v \Delta v_{\max}, 0, v_{\max}^A)
\end{equation*}
The heading angle $\theta_A^t$ is modified by $\lambda_\theta$, with a maximum angular change of $\Delta\theta_{\max}$ radians per time step
\begin{equation*}
    \theta_A^{t+1} = \theta_A^t + \lambda_\theta \Delta\theta_{\max}
\end{equation*}
The desired velocity is defined by
\begin{equation}
\label{eq:desired_velocity}
    v_{des} = [v_Acos\theta_A, v_Asin\theta_A]
\end{equation}

\subsubsection{Reward Function}
The reward structure consists of base rewards for the fundamental objectives, enhanced with potential-based shaping rewards to accelerate convergence while preserving optimality. The immediate reward at time $t$ is defined as

\begin{equation*}
r_t = r_{\text{base}} + r_{\text{shaping}}
\end{equation*}
Base reward is composed of the main objectives. Here, three main objectives are considered. Termination reward to provide feedback for navigation success.
\begin{equation*}
r_T = \begin{cases}
+k_T & \text{if success,}\\
0 & \text{otherwise.}
\end{cases}
\end{equation*}
Collision penalty.
\begin{equation*}
r_{\text{collision}} = \begin{cases}
-k_C & \text{if collision detected,}\\
0 & \text{otherwise.}
\end{cases}
\end{equation*}
Velocity reward to maintain velocity above threshold for faster navigation:
\begin{equation*}
r_V = k_V(v_A - v_c)
\end{equation*}
To accelerate learning three shaping rewards~\eqref{eq:shaping_rewards} are used using potential functions $\Phi_D(s) = -k_D \cdot d$, $\Phi_\Theta(s) = -k_\Theta \cdot |\alpha|$, and $\Phi_{\text{obs}}(s) = k_{\text{obs}} \sum_{i=1}^{9} d_i$.
\begin{equation*}
r_D = k_D(d_p - d_c)
\end{equation*}
\begin{equation*}
r_\Theta = k_\Theta(|\alpha_p| - |\alpha_c|)\pi
\end{equation*}
\begin{equation*}
r_{\text{obs}} = k_{\text{obs}}\sum_{i=1}^{9}(d^c_i - d^p_i)
\end{equation*}

where $d_c$ is the current distance to target, $d_p$ is the previous distance, $\alpha_c$ is the current alignment angle, $\alpha_p$ is the previous alignment angle, $d^c_i$ is the current distance along the $i$-th ray, and $d^p_i$ is the previous distance. As the neural networks are approximators and $\gamma$ in~\eqref{eq:shaping_rewards} is a hyperparameter which is almost equal to unity, this term is neglected in the shaping reward functions.

\subsection{Agent Training}

\begin{figure}[h!]
    \centering
    \includegraphics[width=\columnwidth]{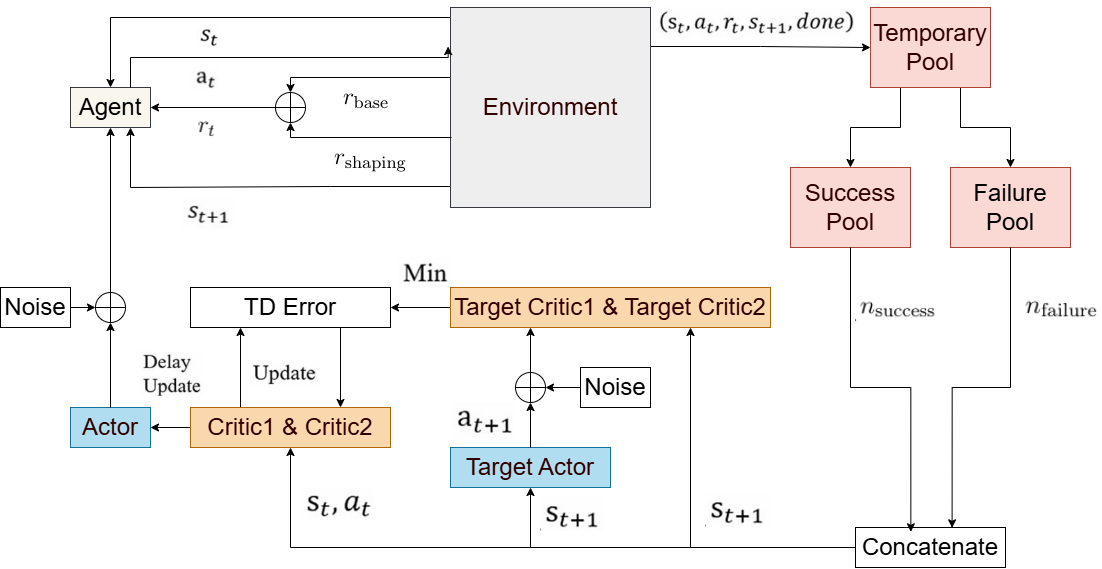}
    \caption{The MPTD3 Architecture.}
    \label{fig:algorithm}
\end{figure}

The MPTD3 algorithm is implemented based on the Stable Baselines 3 (SB3)\cite{raffin2021stable} TD3 implementation. The standard TD3 \cite{fujimoto2018addressing} architecture is extended by introducing a multi-buffer experience replay system to better manage successful and unsuccessful trajectories. The block diagram of the training architecture is illustrated in Fig.~\ref{fig:algorithm}.

As shown in Fig.~\ref{fig:algorithm} MPTD3 maintains three distinct replay pools: a temporary pool(a FIFO deque) for storing the most recent experiences, a success pool for storing experiences from successful episodes and for experiences that overflow the temporary pool and a failure pool for storing experiences from unsuccessful trajectories. All new experiences are first added to the temporary pool. When the temporary pool overflows, the oldest experience is removed and unconditionally added to the success pool. At the end of each episode, all the experiences from the temporary pool are reallocated to either the success or failure pool depending on the final outcome.

Consider a sampling configuration where the success-to-total ratio is denoted by $\eta$, and the batch size is represented by $B$. Let $D_{\text{success}}$ and $D_{\text{failure}}$ be the current number of experiences in the success and failure buffers respectively. The allocation strategy for experience sampling from each repository is determined as follows:

\begin{equation*}
\begin{cases}
n_{\text{failure}} = \min(D_{\text{failure}}, B - \min(D_{\text{success}}, \eta B)) \\
\\
n_{\text{success}} = \min(D_{\text{success}}, B - n_{\text{failure}})
\end{cases}
\end{equation*}

Where, $n_{\text{success}}$ and $n_{\text{failure}}$ represents the quantity of experiences sampled from the success and failure buffer respectively. This adaptive sampling mechanism  prevents convergence in a local minima.

\subsection{CLF-CBF-QP Filter}
\begin{figure}[h!]
    \centering
    \includegraphics[width=\columnwidth]{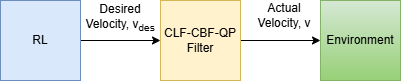}
    \caption{Integration of the CLF-CBF-QP filter with RL.}
    \label{fig:filter}
\end{figure}

To ensure safe, stable and zero-shot transfer capability, a suitable CLF-CBF based Quadratic Programming (QP) filter is proposed, augmented on top of the RL policy. The RL agent proposes a desired velocity vector $v_{\text{des}} \in \mathbb{R}^2$~\eqref{eq:desired_velocity}, which is subsequently modified by the QP filter before being applied to the UAV to guarantee stability, safety and zero-shot transfer ability. The block diagram of the filter integration is shown in Fig.~\ref{fig:filter}

To ensure the UAV continuously converges toward the destination, a Control Lyapunov Function (CLF) is defined based on the Euclidean distance between the UAV position $p_A \in \mathbb{R}^2$ and the destination position $p_T \in \mathbb{R}^2$
\begin{equation}
\label{eq:tracking_clf}
    V(x) = \frac{1}{2} \| p_A - p_T \|^2
\end{equation}
Taking the time derivative yields $\dot{V}(x) = (p_A - p_T)^T (v - v_T)$, where $v$ is the applied UAV velocity and $v_T = \dot{p}_T$ is the velocity of the target. 
Isolating the term containing $v$ yields from the CLF derivative constraint~\eqref{eq:clf_condition} yields
\begin{equation}
\label{eq:clf_constraints}
    (p_A - p_T)^T v \leq (p_A - p_T)^T v_T - \gamma_{\text{clf}} (V(x))
\end{equation}
Where $\gamma_{\text{clf}}$ is a class $\mathcal{K}$ function. Assuming a physical maximum velocity bound for the UAV of $\|v\| \leq v_{\text{max}}$, the minimum possible value of the left-hand side is $-\|p_A - p_T\|v_{\text{max}}$. Substituting this ensures that strict CLF descent remains physically possible
\begin{equation*}
    -\|p_A - p_T\|v_{\text{max}} \leq (p_A - p_T)^T v_T - \gamma_{\text{clf}} (V(x))
\end{equation*}
So. sufficient condition for CLF feasibility
\begin{equation}
    \gamma_{\text{clf}} (V(x)) \leq (p_A - p_T)^T v_T + \|p_A - p_T\|v_{\text{max}}
\end{equation}
For the worst-case scenario, $(p_A - p_T)^T v_T$ reaches its minimum value of 
$-\|p_A - p_T\| v_{T,\max}$, assuming $v_{T,\max}$ is the maximum velocity of 
the target. Substituting this into the inequality yields
\begin{equation*}
    \gamma_{\mathrm{clf}}(V(x)) \leq \|p_A - p_T\| (v_{\max} - v_{T,\max})
\end{equation*}

So, a robust class $\mathcal{K}$ function that satisfies the mathematical constraints
\begin{equation}
    \gamma_{\mathrm{clf}}(V(x)) = k_{\mathrm{clf}}\sqrt{V(x)}
\end{equation}
Where
\begin{equation*}
    0 < k_{\mathrm{clf}} \leq \sqrt{2} (v_{\max} - \|v_T\|_{\max})
\end{equation*}

To guarantee collision-free navigation in the presence of obstacles, the safe set is defined by estimating a dynamic bounding geometry using an onboard sensor that returns the distance observed from 24 rays covering $360^{\circ}$. Let $Q_{\text{active}}$ be the set of LiDAR point cloud returns that fall within a predefined activation distance. The detected obstacle geometry is bounded by a circle whose centroid $c \in \mathbb{R}^2$ is estimated as the mean of the active points
\begin{equation*}
    c = \frac{1}{|Q_{\text{active}}|} \sum_{q \in Q_{\text{active}}} q
\end{equation*}
The observed radius is defined as $R_{\text{obs}} = \max_{q} \|q - c\|$. The effective safety radius is $R = R_{\text{obs}} + d_{\text{safe}}$, where $d_{\text{safe}}$ is the minimum allowable margin. The Control Barrier Function for obstacle avoidance is formulated as
\begin{equation}
\label{eq:obstacle_cbf}
    h_{\text{obs}}(x) = \| p_A - c \|^2 - R^2
\end{equation}
Taking the time derivative yields $\dot{h}_{\text{obs}}(x) = 2(p_A - c)^T v - 2(p_A - c)^T \dot{c} -  2R\dot{R}$. Isolating the term containing the control input $v$ from the obstacle CBF derivative constraint~\eqref{cbf_condition} yields
\begin{equation}
\label{eq:cbf_constraints}
    2(p_A - c)^T v \geq 2(p_A - c)^T \dot{c} + 2R\dot{R} - \alpha_{\text{ocbf}} (h_{\text{obs}}(x))
\end{equation}
Where $\alpha_{\text{ocbf}}$ is a extended class $\mathcal{K}_\infty$ function.
The maximum possible value of the left-hand side is $2\|p_A - c\|v_{\text{max}}$. For the constraint inequality to remain feasible, this maximum achievable value must be greater than or equal to the right-hand side
\begin{equation*}
    2\|p_A - c\|v_{\text{max}} \geq 2(p_A - c)^T \dot{c} + 2R\dot{R} - \alpha_{\text{ocbf}} (h_{\text{obs}}(x))
\end{equation*}
So, the sufficient condition for obstacle CBF feasibility
\begin{equation}
    \alpha_{\text{ocbf}} (h_{\text{obs}}(x)) \geq 2(p_A - c)^T \dot{c} + 2R\dot{R} - 2\|p_A - c\|v_{\text{max}}
\end{equation}
Let $\dot{c}_{\max}$ be the maximum speed of the obstacle's centroid, and $\dot{R}_{\max}$ be the maximum rate of expansion of the obstacle's bounding radius. The maximum value of the term $2(p_A - c)^T \dot{c}$ is $2\|p_A - c\| \dot{c}_{\max}$. Therefore, the right-hand side is strictly bounded by $2\|p_A - c\| ( \dot{c}_{\max} - v_{\max} ) + 2R\dot{R}_{\max}$

Because an extended class $\mathcal{K}_{\infty}$ function must strictly satisfy $\alpha(0) = 0$. Therefore, the system must satisfy
\begin{equation} 
    v_{\max} \geq \dot{c}_{\max} + \dot{R}_{\max}
\label{cbfsufficient}
\end{equation}
As long as the UAV's physical speed limit complies with the above necessary condition, a standard linear function is a perfect candidate for $\alpha_{\mathrm{ocbf}}$.
\begin{equation}
    \alpha_{\mathrm{ocbf}}(h_{\mathrm{obs}}(x)) = k_{\mathrm{cbf}} h_{\mathrm{obs}}(x)
\end{equation}
Where $k_{\mathrm{cbf}} > 0$.

To restrict the UAV within the designated workspace, additional linear barrier functions are defined for the four boundaries with a safety margin $m$
\begin{equation}
\label{eq:boundary_constraints}
\begin{aligned}
    h_{\text{left}}(x) &= x - m \\
    h_{\text{right}}(x) &= (w - m) - x \\
    h_{\text{bottom}}(y) &= y - m \\
    h_{\text{top}}(y) &= (h - m) - y
\end{aligned}
\end{equation}
The derivative constraints of~\eqref{eq:boundary_constraints} can be derived in the same manner as~\eqref{eq:cbf_constraints}.

The proposed CLF-CBF-QP filter incorporates both the CLF~\eqref{eq:clf_controller} and CBF~\eqref{eq:cbf_controller} controller utilizing the proposed tracking CLF ~\eqref{eq:tracking_clf}, obstacle avoidance CBF~\eqref{eq:obstacle_cbf}, and boundary restriction CBFs~\eqref{eq:boundary_constraints}. The filter minimizes the squared Euclidean distance between the actual applied velocity $v$ and the desired velocity $v_{\text{des}}$~\eqref{eq:desired_velocity}. To ensure the QP remains feasible, slack variables $\delta_1$ and $\delta_2$ are introduced to soften the tracking CLF derivative constraint~\eqref{eq:clf_constraints} and the derivative constraints of the boundary restriction CBFs~\eqref{eq:boundary_constraints}, respectively. The derivative constraint~\eqref{eq:cbf_constraints} of obstacle avoidance CBF is enforced as a strict hard constraint. The proposed optimization problem is formulated as
\begin{equation}
\label{eq:optimaization}
\begin{aligned}
    \min_{\substack{v \in \mathbb{R}^2 \\ \delta_1, \delta_2 \in \mathbb{R}}} \quad & \frac{1}{2} \| v - v_{\text{des}} \|^2 + \frac{\lambda_{\text{slack}}}{2} (\delta_1^2 + \delta_2^2) \\
    \text{s.t.} \quad 
    & (p_A - p_T)^T (v - v_T) + \gamma_{\text{clf}} (V(x)) \leq \delta_1 \\
    & 2(p_A - c)^T (v - \dot{c}) - 2R\dot{R} + \alpha_{\text{ocbf}} (h_{\text{obs}}(x)) \geq 0 \\
    & v_x + \alpha_{\text{bcbf}} h_{\text{left}}(x) \geq \delta_2 \\
    & -v_x + \alpha_{\text{bcbf}} h_{\text{right}}(x) \geq \delta_2 \\
    & v_y + \alpha_{\text{bcbf}} h_{\text{bottom}}(x) \geq \delta_2 \\
    & -v_y + \alpha_{\text{bcbf}} h_{\text{top}}(x) \geq \delta_2 \\
    & \|v\| \leq v_{\text{max}}
\end{aligned}
\end{equation}

where $\alpha_{\text{bcbf}} > 0$ defines the strictness of the boundary barrier response, and $\lambda_{\text{slack}}$ applies a heavy penalty on violation of the tracking CLF and boundary restriction CBFs. To summarize the deployment phase, Algorithm \ref{alg:navigation_loop} details the step-by-step process where the RL agent's proposed velocity is formally verified and, if necessary, modified by the optimization filter to guarantee zero-shot safety prior to application.

\begin{algorithm}[h!]
\caption{Zero-Shot Safe Navigation via RL and CLF-CBF-QP Filter}
\label{alg:navigation_loop}
\begin{algorithmic}[1]
\REQUIRE Pre-trained MPTD3 policy $\pi_\theta$, Initial UAV state $x_0$, Destination $p_T$, Control limits $v_{\max}$, Safety margin $d_{\text{safe}}$
\STATE Initialize state $x \leftarrow x_0$
\WHILE{$\|p_A - p_T\| > \text{success\_threshold}$}
    \STATE Obtain normalized observation $s_t$ from sensors
    \STATE Query policy for continuous action: $[\lambda_v, \lambda_\theta] \leftarrow \pi_\theta(s_t)$
    \STATE Calculate desired velocity $v_{\text{des}}$ based on $v_A, \theta_A$, and $[\lambda_v, \lambda_\theta]$ \COMMENT{Eq. \ref{eq:desired_velocity}}
    \STATE Compute tracking CLF $V(x) = \frac{1}{2}\|p_A - p_T\|^2$ \COMMENT{Eq. \ref{eq:tracking_clf}}
    \STATE Obtain LiDAR point cloud and filter active returns $Q_{\text{active}}$
    \IF{$|Q_{\text{active}}| > 0$}
        \STATE Estimate obstacle centroid $c$ and bounding radius $R_{\text{obs}}$
        \STATE Set effective safety radius $R \leftarrow R_{\text{obs}} + d_{\text{safe}}$
        \STATE Evaluate obstacle CBF $h_{\text{obs}}(x) = \|p_A - c\|^2 - R^2$ \COMMENT{Eq. \ref{eq:obstacle_cbf}}
    \ENDIF
    \STATE Evaluate boundary CBFs $h_{\text{left}}, h_{\text{right}}, h_{\text{bottom}}, h_{\text{top}}$ \COMMENT{Eq. \ref{eq:boundary_constraints}}
    \STATE Solve the CLF-CBF-QP optimization problem~\eqref{eq:optimaization}.
    \STATE Retrieve optimal, safe velocity vector $v^*$
    \STATE Apply $v^*$ to UAV dynamics and update state $x$
\ENDWHILE
\end{algorithmic}
\end{algorithm}

\section{Results}
The implementation utilizes a combination of established Python libraries such as Stable-Baselines3 2.7.0~\cite{raffin2021stable}, Gymnasium 1.1.1~\cite{towers2024gymnasium}, PyBullet 3.21~\cite{panerati2021learning}.

To evaluate the performance of the RL algorithm, an environment with $1000 \times 1000$ area, 20 static and 5 dynamic obstacles is created. The position of the UAV, the destination point and the obstacles' positions are initialized randomly on the entire area at the beginning of each episode. Training of RL is conducted for $10^6$ timesteps, with each episode capped at 1000 steps to prevent excessively long episodes. After training the RL model is tested in the same environment for 1000 episodes. 

The effectiveness of randomly initializing obstacles position during training is validated by the UAV's performance as shown in Table~\ref{table:fixedvsrandom_results}. It is observed that the random initialization increased the success rate by almost 19\% and reduced the average episode steps by approximately 32\%.

\begin{table}[h!]
\renewcommand{\arraystretch}{1.3} 
\caption{Performance during test for UAV trained with fixed vs random obstacle position}
\label{table:fixedvsrandom_results}
\centering
\begin{tabular}{|c|c|c|}
\hline
\textbf{Type} & \textbf{Success} & \textbf{Average Episode Steps} \\ \hline
Fixed & 774 & 396.53 \\ \hline
Random & 962 & 267.29 \\ \hline
\end{tabular}
\end{table}

To evaluate the impact of the proposed potential-based shaping reward, a comparative analysis is conducted against the reward model from \cite{xu2022autonomous}, as the authors rewarded the UAV for success, punished for failure and low velocity, and designed the other rewards from heuristic. Table~\ref{table:prbs_results} shows that the success rate increased slightly but the average steps is reduced by almost 19\% both during training and test. Thus it empirically proves that PRBS preserves the optimality better than the heuristics.

\begin{table}[h!]
\renewcommand{\arraystretch}{1.3} 
\caption{Performance of Proposed Reward vs Baseline Reward} 
\label{table:prbs_results}
\centering
\begin{tabular}{|c|c|c|c|c|}
\hline
\multirow{2}{*}{\textbf{Type}} & \multicolumn{2}{c|}{\textbf{Success}} & \multicolumn{2}{c|}{\textbf{Average Episode Steps}} \\ \cline{2-5}
 & \textbf{Train} & \textbf{Test} & \textbf{Train} & \textbf{Test} \\ \hline
\textbf{Baseline} & 91.07\% & 95.6\% & 336.78 & 328.89 \\ \hline
\textbf{Proposed} & 91.58\% & 96.2\% & 270.3 & 267.29 \\ \hline
\end{tabular}
\end{table}

To validate the safety, stability and zero-shot transfer performance of the CLF-CBF-QP filter, two additional $100 \times 100$ environments with 40 static and 10 dynamic obstacles are created. In one environment the target is static and dynamic in other. The test results shown for these environments are for 100 episodes. For the static target navigation, the UAV is initialized at the bottom of y axis, the destination point is at the top and the obstacles are randomly scattered at the middle. And for the dynamic target tracking, the obstacles are initialized randomly in $80 \times 80$ area in the middle. The target moves in this $80 \times 80$ area with a constant velocity magnitude but changes its direction with 10\% probability. The performance of the UAV, including an ablation study across increasing functionalities, is given in Table \ref{table:env2_results} and Table \ref{table:env3_results}. It is observed from the results that the filter serves as a stable and safe guide for the UAV. The pretrained RL agent can be used in complex scenarios without further training with outstanding performance. The performance without the velocity constraint is also observed to validate \eqref{cbfsufficient}.

\begin{table}[htbp]
\centering
\caption{Performance Comparison: Static Target}
\label{table:env2_results}
\setlength{\tabcolsep}{4pt} 
\small 
\begin{tabular}{@{}lccccc@{}}
\toprule
\textbf{Method} & \textbf{Succ.} & \textbf{Coll.} & \textbf{O.o.B.} & \textbf{T.O.} & \textbf{Steps} \\ \midrule
RL(1 Hz)               & 44             & 2              & 51              & 3             & 102.84          \\
RL + Decision(10 Hz)           & 73             & 4              & 22              & 1             & 74.95           \\
RL + CLF-QP(10 Hz)              & 75             & 24             & 1               & 0             & 52.93           \\
RL + CLF-CBF-QP(10 Hz)          & 99   & 1     & 0      & 0    & 59.52           \\ 
\makecell[l]{RL + CLF-CBF-QP(10 Hz) \\ without velocity constraint} & 100  & 0  & 0   & 0  & 59.55  \\           \\ 
\bottomrule
\end{tabular}
\end{table}

\begin{table}[htbp]
\centering
\caption{Performance Comparison: Dynamic Target}
\label{table:env3_results}
\setlength{\tabcolsep}{4pt} 
\small 
\begin{tabular}{@{}lccccc@{}}
\toprule
\textbf{Method} & \textbf{Succ.} & \textbf{Coll.} & \textbf{O.o.B.} & \textbf{T.O.} & \textbf{Steps} \\ \midrule
RL(1 Hz)                & 65             & 4              & 14              & 17            & 157.25          \\
RL + Decision(10 Hz)            & 79             & 3              & 10              & 8             & 118.85          \\
RL + CLF-QP(10 Hz)              & 59             & 38             & 3               & 0             & 45.88           \\
RL + CLF-CBF-QP(10 Hz)          & 98   & 2     & 0      & 0    & 55.52           \\ 
\makecell[l]{RL + CLF-CBF-QP(10 Hz) \\ without velocity constraint} & 100   & 0     & 0      & 0    & 55.91           \\ 
\bottomrule
\end{tabular}
\end{table}

\section{Conclusion}

This study introduces a generalized framework that utilizes the model free training feature of RL to establish a general controller. Potential Based Reward Functions are used to preserve the desired optimality of the navigation problem. Moreover, a CLF-CBF-QP filter is augmented on top of RL to guide the controller in complex environments. The approach is empirically validated in simulated environments and the obtained results shows significant improvements. A relevant direction of future work is to utilized the proposed framework in a practical real world environment and finding suitable Lyapunov and Barrier functions for specific scenarios.

\bibliographystyle{IEEEtran}
\bibliography{references}

\begin{thebibliography}{10}
\providecommand{\url}[1]{#1}
\csname url@samestyle\endcsname
\providecommand{\newblock}{\relax}
\providecommand{\bibinfo}[2]{#2}
\providecommand{\BIBentrySTDinterwordspacing}{\spaceskip=0pt\relax}
\providecommand{\BIBentryALTinterwordstretchfactor}{4}
\providecommand{\BIBentryALTinterwordspacing}{\spaceskip=\fontdimen2\font plus
\BIBentryALTinterwordstretchfactor\fontdimen3\font minus \fontdimen4\font\relax}
\providecommand{\BIBforeignlanguage}[2]{{%
\expandafter\ifx\csname l@#1\endcsname\relax
\typeout{** WARNING: IEEEtran.bst: No hyphenation pattern has been}%
\typeout{** loaded for the language `#1'. Using the pattern for}%
\typeout{** the default language instead.}%
\else
\language=\csname l@#1\endcsname
\fi
#2}}
\providecommand{\BIBdecl}{\relax}
\BIBdecl

\bibitem{li2026review}
Y.~Li, G.~Deng, H.~Zhao, B.~Liu, C.~Liu, W.~Qian, and X.~Qiao, ``A review of uav remote sensing technology applications in common gramineous crops,'' \emph{Information Processing in Agriculture}, 2026.

\bibitem{ji2025uav}
Y.~Ji, K.~Song, H.~Wen, X.~Xue, Y.~Yan, and Q.~Meng, ``Uav applications in intelligent traffic: Rgbt image feature registration and complementary perception,'' \emph{Advanced Engineering Informatics}, vol.~63, p. 102953, 2025.

\bibitem{zhang2026uav}
L.~Zhang, Y.~Wang, X.~Xue, W.~Huang, T.~Yang, H.~Zhu, and Y.~Lan, ``Uav remote sensing-driven precision variable management in cotton: technological framework, applications, and research outlook,'' \emph{Computers and Electronics in Agriculture}, vol. 243, p. 111426, 2026.

\bibitem{kendoul2012survey}
F.~Kendoul, ``Survey of advances in guidance, navigation, and control of unmanned rotorcraft systems,'' \emph{Journal of Field Robotics}, vol.~29, no.~2, pp. 315--378, 2012.

\bibitem{rezwan2022artificial}
S.~Rezwan and W.~Choi, ``Artificial intelligence approaches for uav navigation: Recent advances and future challenges,'' \emph{IEEE access}, vol.~10, pp. 26\,320--26\,339, 2022.

\bibitem{aburaya2024review}
A.~Aburaya, H.~Selamat, and M.~T. Muslim, ``Review of vision-based reinforcement learning for drone navigation: A. aburaya et al.'' \emph{International Journal of Intelligent Robotics and Applications}, vol.~8, no.~4, pp. 974--992, 2024.

\bibitem{xu2022autonomous}
G.~Xu, W.~Jiang, Z.~Wang, and Y.~Wang, ``Autonomous obstacle avoidance and target tracking of uav based on deep reinforcement learning,'' \emph{Journal of Intelligent \& Robotic Systems}, vol. 104, no.~4, p.~60, 2022.

\bibitem{jiang2024autonomous}
W.~Jiang, T.~Cai, G.~Xu, and Y.~Wang, ``Autonomous obstacle avoidance and target tracking of uav: Transformer for observation sequence in reinforcement learning,'' \emph{Knowledge-Based Systems}, vol. 290, p. 111604, 2024.

\bibitem{ma2023target}
B.~Ma, Z.~Liu, W.~Zhao, J.~Yuan, H.~Long, X.~Wang, and Z.~Yuan, ``Target tracking control of uav through deep reinforcement learning,'' \emph{IEEE Transactions on Intelligent Transportation Systems}, vol.~24, no.~6, pp. 5983--6000, 2023.

\bibitem{chen2026learning}
Z.~Chen and J.~Xuan, ``Learning unknown reward function for drone navigation based on inverse deep reinforcement learning,'' \emph{Neural Computing and Applications}, vol.~38, no.~4, p.~51, 2026.

\bibitem{wang2020deep}
C.~Wang, J.~Wang, J.~Wang, and X.~Zhang, ``Deep-reinforcement-learning-based autonomous uav navigation with sparse rewards,'' \emph{IEEE Internet of Things Journal}, vol.~7, no.~7, pp. 6180--6190, 2020.

\bibitem{chansuparp2022novel}
M.~Chansuparp and K.~Jitkajornwanich, ``A novel augmentative backward reward function with deep reinforcement learning for autonomous uav navigation,'' \emph{Applied Artificial Intelligence}, vol.~36, no.~1, p. 2084473, 2022.

\bibitem{tsourveloudis2025uav}
C.~Tsourveloudis and L.~Doitsidis, ``Uav navigation using reinforcement learning: A systematic approach to progressive reward function design,'' 2025.

\bibitem{ng1999policy}
A.~Y. Ng, D.~Harada, and S.~Russell, ``Policy invariance under reward transformations: Theory and application to reward shaping,'' in \emph{Icml}, vol.~99.\hskip 1em plus 0.5em minus 0.4em\relax Citeseer, 1999, pp. 278--287.

\bibitem{peng2023design}
C.~Peng, X.~Liu, and J.~Ma, ``Design of safe optimal guidance with obstacle avoidance using control barrier function-based actor--critic reinforcement learning,'' \emph{IEEE Transactions on Systems, Man, and Cybernetics: Systems}, vol.~53, no.~11, pp. 6861--6873, 2023.

\bibitem{marzari2025designing}
L.~Marzari, F.~Trotti, E.~Marchesini, and A.~Farinelli, ``Designing control barrier function via probabilistic enumeration for safe reinforcement learning navigation,'' \emph{IEEE Robotics and Automation Letters}, 2025.

\bibitem{chen2025collision}
H.~Chen, F.~Zhang, and B.~Aksun-Guvenc, ``Collision avoidance in autonomous vehicles using the control lyapunov function--control barrier function--quadratic programming approach with deep reinforcement learning decision-making,'' \emph{Electronics}, vol.~14, no.~3, p. 557, 2025.

\bibitem{emam2022safe}
Y.~Emam, G.~Notomista, P.~Glotfelter, Z.~Kira, and M.~Egerstedt, ``Safe reinforcement learning using robust control barrier functions,'' \emph{IEEE Robotics and Automation Letters}, vol.~10, no.~3, pp. 2886--2893, 2022.

\bibitem{xia2025multi}
H.~Xia, Q.~Qi, X.~Yang, X.~Ju, and H.~Su, ``Multi-uav-ugv collision-free tracking control via control barrier function-based reinforcement learning,'' in \emph{2025 IEEE/RSJ International Conference on Intelligent Robots and Systems (IROS)}.\hskip 1em plus 0.5em minus 0.4em\relax IEEE, 2025, pp. 17\,115--17\,121.

\bibitem{slotine1991applied}
J.-J.~E. Slotine, W.~Li \emph{et~al.}, \emph{Applied nonlinear control}.\hskip 1em plus 0.5em minus 0.4em\relax Prentice hall Englewood Cliffs, NJ, 1991, vol. 199, no.~1.

\bibitem{ames2019control}
A.~D. Ames, S.~Coogan, M.~Egerstedt, G.~Notomista, K.~Sreenath, and P.~Tabuada, ``Control barrier functions: Theory and applications,'' in \emph{2019 18th European control conference (ECC)}.\hskip 1em plus 0.5em minus 0.4em\relax Ieee, 2019, pp. 3420--3431.

\bibitem{raffin2021stable}
A.~Raffin, A.~Hill, A.~Gleave, A.~Kanervisto, M.~Ernestus, and N.~Dormann, ``Stable-baselines3: Reliable reinforcement learning implementations,'' \emph{Journal of machine learning research}, vol.~22, no. 268, pp. 1--8, 2021.

\bibitem{fujimoto2018addressing}
S.~Fujimoto, H.~Hoof, and D.~Meger, ``Addressing function approximation error in actor-critic methods,'' in \emph{International conference on machine learning}.\hskip 1em plus 0.5em minus 0.4em\relax PMLR, 2018, pp. 1587--1596.

\bibitem{towers2024gymnasium}
M.~Towers, A.~Kwiatkowski, J.~Terry, J.~U. Balis, G.~De~Cola, T.~Deleu, M.~Goul{\~a}o, A.~Kallinteris, M.~Krimmel, A.~KG \emph{et~al.}, ``Gymnasium: A standard interface for reinforcement learning environments,'' \emph{arXiv preprint arXiv:2407.17032}, 2024.

\bibitem{panerati2021learning}
J.~Panerati, H.~Zheng, S.~Zhou, J.~Xu, A.~Prorok, and A.~P. Schoellig, ``Learning to fly—a gym environment with pybullet physics for reinforcement learning of multi-agent quadcopter control,'' in \emph{2021 IEEE/RSJ International Conference on Intelligent Robots and Systems (IROS)}.\hskip 1em plus 0.5em minus 0.4em\relax IEEE, 2021, pp. 7512--7519.

\end{thebibliography}

\end{document}